

\magnification=\magstep1
\baselineskip=12pt
\font\ab=cmr9
\overfullrule=0pt
\font\twelvebf=cmbx12
\hsize = 5 in
\vsize = 7 in
\vglue .7 in
\centerline{\twelvebf  On the Squeezed Vacuum States}
\centerline{\twelvebf  Corresponding to the General}
\centerline{\twelvebf  Two-Mode Bogolubov Transformation}
\vskip .7 in
\centerline{\bf Oleg Mokhov\footnote{$^{\dag}$}
{On leave of absence from
{\it VNIIFTRI, Mendeleevo, Moscow Region 141570, Russia}}}
\smallskip
\centerline{\it Department of Mathematics}
\centerline{\it Bilkent University}
\centerline{\it 06533 Bilkent, Ankara, Turkey}
\vskip 1in
\leftskip = 30 pt
\rightskip = 30 pt
\centerline{\bf Abstract}
\smallskip

{\ab In this paper the explicit form of the operator of transformation
of the vacuum states for the general two-mode Bogolubov transformation
is found}
\par
\vfill\eject
\hsize = 5 in
\leftskip = 0 pt
\rightskip = 0 pt
\vskip .5in

\noindent{\bf 1  Introduction}
\vskip .2in

A number of problems of quantum optics is connected with investigation
of model Hamiltonians given by Hermitian quadratic form of Bose operators:
$$H = {1 \over 2} \sum_{k,l} [A_{kl}a_k^{\dagger} a_l^{\dagger} +
A_{kl}^*a_k a_l + 2 B_{kl} a_k^{\dagger} a_l],  \eqno{(1)}$$
$$A_{kl} =A_{lk},\
B_{kl} = B_{lk}^*,$$
$$[a_k,a_l^{\dagger}] = \delta_{kl},\   [a_k,a_l] =
[a_k^{\dagger},a_l^{\dagger}$$
For diagonalization of such quadratic Hamiltonians the canonical
Bogolubov transformations [1] are used:
$$b_k = \sum_m (u_{km} a_m + v_{km} a_m^{\dagger}),  \eqno {(2)}$$
$$b_k^{\dagger} = \sum_m (v_{km}^* a_m + u_{km}^* a_m^{\dagger}),$$
i.e. the linear transformations of Bose operators, preserving the
commutational relations of the algebra of Bose operators:
$$[b_k,b_l^{\dagger}] = \delta_{kl},\   [b_k,b_l] =
[b_k^{\dagger},b_l^{\dagger}
\eqno {(3)}$$
In particular, the use of the general canonical Bogolubov transformations
is necessary for the study of non-classical optical effects such as squeezing
[2,3] (see also the survey [4]). In the case of a single mode, for the
Hermitian
quadrature operators
$$Q={(a^{\dagger} + a) \over 2},\   P=i {(a^{\dagger} - a) \over 2},
\  [Q,P] = {i \over 2}  \eqno{(4)}$$
we have the Heisenberg uncertainty relation for quantum fluctuations:
$$<(\Delta Q)^2><((\Delta P)^2> \geq  \left
|{<[Q,P]> \over 4} \right |^2 \equiv {1 \over 16}
 \eqno {(5)}$$
The vacuum and coherent states give
$$<(\Delta Q)^2> = <(\Delta P)^2> = {1 \over 4}, $$
i.e. they realize the minimum-uncertainty states with
$$<(\Delta Q)^2> <(\Delta P)^2> = {1 \over 16} \eqno{(6)}$$
For squeezed states
$$<(\Delta Q)^2> = {1 \over 4} \exp (- 2 r),\
<(\Delta P)^2> = {1 \over 4} \exp (2 r)  \eqno{(7)}$$
and they also realize the minimum-uncertainty states but if $ r
\not =  0$ then one of the qaudrature operators satisfies to
the relation $<(\Delta X)^2> < {1 \over 4} $ ( $X=Q $ or $X=P$),
i.e. the quantum fluctuations in one of the field quadrature
can be below than the usual vacuum level. These squeezed states can be
obtained with the use of the general one-mode canonical Bogolubov
transformation:
$$b = e^{i \phi} \cosh r \   a + e^{i \psi} \sinh r \   a^{\dagger},\
b^{\dagger} = e^{- i \phi} \cosh r \   a^{\dagger} + e^{- i \psi}
\sinh r  \  a  \eqno{(8)}$$
In this connection one needs an explicit form of the operators
transforming the Fock states of Bose-fields $a_k$ to the
corresponding Fock states of Bose-fields $ b_k$ (for
Bose-fields $a_k$ and $b_k$ connected by the Bogolubov
transformation (2),(3)). First of all, it is necessary to know
the operator of transformation for the corresponding vacuum
states $a_k |0>^{(a)} = 0$, $b_k |0>^{(b)} = 0,$ $ 1 \leq k \leq N,$
where $|0>^{(x)} = \prod_{m=1}^N |0>_m^{(x)}.$
For one-mode case (a single Bose-field a) such unitary operator
(so called "squeezed" operator) was found by Stoler [2] for the
general one-mode canonical Bogolubov transformation (8). For two-mode
case analogous "squeezed" operators was constructed only for
Bogolubov transformations of the very special form, as a matter of
fact, only for the direct combination of one-mode Bogolubov
transformations (after some elementary transformations of
Bose-fields) [5,6], see also [4]. Meanwhile, for quantum optics
the general two-mode case is also very important. Particularly,
the model of polariton [7,8] corresponds to the more general
case of two-mode quadratic Hamiltonian. In this paper we attempted
to study the most general two-mode case. The explicit formulas
connecting the general Bogolubov transformation with the corresponding
unitary "squeezed" operator are very complicated but we present here
the simple explicit form of the operator of transformation for
the vacuum states for the most general two-mode Bogolubov
transformation.

Author is very grateful to Prof. A.S.Shumovsky for valuable
discussions on the problems of quantum optics.

\vskip .3 in

\noindent{\bf 2  General scheme for N-mode case  }
\vskip .2 in

We shall consider here the N-mode case of the usual operators
of creation and annihilation for photons:
$$ a_k |n>_k = \sqrt {n} |n-1>_k,\
a_k^{\dagger} |n>_k = \sqrt {n + 1} |n + 1>_k,\
a_k^{\dagger} a_k |n>_k = n |n>_k$$
Let be given an arbitrary unitary transformation of Bose operators
$a$:
$$b_k=Ua_kU^{\dagger}, \eqno{(9)}$$
where $U(a,a^{\dagger})$ is an unitary operator:
$U^{\dagger} =U^{-1}$, $b_k^{\dagger} = Ua_k^{\dagger}U^{\dagger}.$
The transformation (9) preserve the commutational relations (3) for
the algebra of Bose operators. The vacuum state of Bose-fields
$b$ is determined by the following formula:
$$|0>^{(b)} = U|0>^{(a)}$$
For arbitrary Fock states we have
$$\prod_{k=1}^N |n_k>_k^{(b)} = \prod_{k=1}^N
{(b_k^{\dagger})^{n_k} \over \sqrt {(n_k)!}} |0>_k^{(b)}=$$
$$U \prod_{k=1}^N {(a_k^{\dagger})^{n_k} \over \sqrt {(n_k)!}
} |0>_k^{(a)} = U \prod_{k=1}^N |n_k>_k^{(a)},$$
i.e. the unitary operator $U$ is the operator of transformation for
all Fock states after the unitary transformation (9) of Bose operators.

Consider now an arbitrary Bogolubov transformation (2),(3). For the
vacuum state of Bose-fields $b$ we have:
$$|0>^{(b)} = \sum_{i_s=0}^{\infty} \lambda_{i_1 \dots i_N}
\prod_{k=1}^N |i_k>_k^{(a)},  \eqno{(10)}$$
$$1 \leq s \leq N$$
The condition
$$b_k |0>^{(b)} =0 $$
gives recurrence formulas for the coefficients $\lambda_{i_1 \dots i_N} $:
$$ \sum_{m=1}^N (u_{km} \lambda_{i_1 \dots (i_m +1) \dots i_N}
\sqrt {i_m +1} + v_{km} \lambda_{i_1 \dots (i_m - 1) \dots i_N}
\sqrt {i_m}) = 0 \eqno{(11)}$$
For $N=1$ and $N=2$ these equations for $\lambda_{i_1 \dots i_N}$
are completely solved in this paper for the corresponding general
Bogolubov transformations (of course, the case $N=1$ is very simple).
After that if we know the operator $S_0(a^{\dagger})$ such that
$$|0>^{(b)} = S_0(a^{\dagger}) |0>^{(a)}$$
we can determine uniquely the corresponding unitary "squeezed"
operator U by recurrence procedure from the necessary relations
for all photon number Fock states:
$$\prod_{k=1}^N |n_k>_k^{(b)} = U \prod_{k=1}^N |n_k>_k^{(a)}$$
\vskip .3 in

\noindent{\bf 3   One-mode case}
\vskip .2in

First of all consider the simple one-mode case of the
general Bogolubov transformation:
$$b= ua + va^{\dagger},\  b^{\dagger} = v^* a + u^* a^{\dagger}$$
The condition $[b,b^{\dagger}] =1$ gives $uu^* -vv^* =1$, i.e.
$$u = e^{i \phi} \cosh r, \  v= e^{i \psi} \sinh r$$
For the vacuum state of Bose-field $b$ we have:
$$|0>^{(b)} = \sum_{i=0}^{\infty} \lambda_i |i>^{(a)}, \eqno {(12)}$$
and from the condition $b|0>^{(b)}=0$ we receive the simple recurrence
formulas for $\lambda_i$:
$$\lambda_{i+1} u \sqrt {i+1} + \lambda_{i-1} v \sqrt {i} =0 \eqno{(13)}$$
It is easy to obtain the general solution of these recurrence
equations (13):
$$\lambda_{2k + 1} =0, \  \lambda_{2k} =(-1)^k \left ({v \over u}\right )^k
{\sqrt {(2k-1)!!} \over \sqrt {(2k)!!}} \lambda_0  \eqno{(14)}$$
Thus,
$$|0>^{(b)} = \sum_{k=0}^{\infty} (-1)^k \left ({v \over u}\right )^k
{\sqrt {(2k-1)!!} \over \sqrt {(2k)!!}} \lambda_0 |2k>^{(a)}
\equiv$$  $$ \left ( \sum_{k=0}^{\infty} (-1)^k \left ({v \over 2u}\right )^k
{(a^{\dagger})^{2k} \over k!} \right ) \lambda_0 |0>^{(a)}, \eqno{(15)}$$
i.e. we receive the following formula for transformation of the
vacuum states in the one-mode case:
$$|0>^{(b)} = \lambda_0 e^{-{v \over 2u} (a^{\dagger})^2 } |0>^{(a)}
\eqno{(16)}$$
The corresponding unitary operator
$$U= \lambda_0 \sum_{k=0}^{\infty} S_k (a^{\dagger}) a^k, \eqno{(17)}$$
where
$$ S_0 (a^{\dagger}) = \exp {\left (- {v \over 2u}
(a^{\dagger})^2\right )}, \eqno{(18)}$$
can be determined uniquely from relations
$$|n>^{(b)} = U |n>^{(a)}$$
which give the reccurence formulas for $S_n(a^{\dagger})$:
$$(v^* a + u^* a^{\dagger})^n S_0 (a^{\dagger}) |0>^{(a)} =
\sum_{k=0}^{\infty} S_k (a^{\dagger}) a^k (a^{\dagger})^n |0>^{(a)}
 \eqno{(19)}$$
For $\phi =0$ we have the well-known unitary "squeezed" operator
of Stoler [2]:
$$U= \exp { [{1 \over 2 } (re^{-i \psi}a^2 - re^{i \psi}(a^{\dagger})^2)]}$$

\vskip .3 in
\noindent{\bf 4   Two-mode case}
\vskip .2 in

Consider the general two-mode Bogolubov transformation:
$$b_1 = u_{11}a_1 + u_{12}a_2 +v_{11}a_1^{\dagger} + v_{12}a_2^{\dagger},$$
$$b_2 = u_{21}a_1 + u_{22}a_2 +v_{21}a_1^{\dagger} + v_{22}a_2^{\dagger},$$
$$b_1^{\dagger} = v_{11}^* a_1 + v_{12}^* a_2 +
u_{11}^* a_1^{\dagger} + u_{12}^* a_2^{\dagger},$$
$$b_2^{\dagger} = v_{21}^* a_1 + v_{22}^* a_2 +
u_{21}^* a_1^{\dagger} + u_{22}^* a_2^{\dagger}, \eqno{(20)}$$
where the coefficients $(u_{ij},v_{ij})$ must satisfy the relations:
$$1)\ \  u_{11}v_{21} + u_{12}v_{22} - u_{21}v_{11} - u_{22}v_{12}=0,$$
$$2)\ \  u_{11}u_{21}^* + u_{12}u_{22}^* -
v_{11}v_{21}^* - v_{12}v_{22}^* = 0,$$
$$3)\ \  u_{11}u_{11}^* + u_{12}u_{12}^* - v_{11}v_{11}^* -
v_{12}v_{12}^* =1,$$
$$4)\ \  u_{21}u_{21}^* +u_{22}u_{22}^* -
v_{21}v_{21}^* - v_{22}v_{22}^* =1,  \eqno {(21)}$$
in order to preserve the commutational relations (3)
of the algebra of Bose operators.
For the vacuum state of Bose-fields $b$ we have the following representation:
$$|0>^{(b)} = \sum_{i,j=0}^{\infty} \lambda_{ij} |i>_1^{(a)}|j>_2^{(a)}
\eqno{(22)}$$
and from the conditions $b_k|0>^{(b)}=0$ we can obtain the recurrence
equations for $\lambda_{ij}$:
$$\lambda_{(i+1)j}u_{11} \sqrt {i+1} + \lambda_{i(j+1)}u_{12}
\sqrt {j+1} + \lambda_{(i-1)j}v_{11} \sqrt {i} + \lambda_{i(j-1)}
v_{12} \sqrt {j} = 0,$$
$$\lambda_{(i+1)j}u_{21} \sqrt {i+1} + \lambda_{i(j+1)} u_{22}
\sqrt {j+1} + \lambda_{(i-1)j}v_{21} \sqrt {i} + \lambda_{i(j-1)}
v_{22} \sqrt {j} =0  \eqno{(23)}$$
Further we shall consider that
$$\Delta_1 \equiv u_{11}u_{22} - u_{12}u_{21} \not = 0$$
The case $\Delta_1 =0$ is very simple and we shall not stop here on it.
Let us introduce also for convenience the following notations:
$$\Delta_2 \equiv u_{11}v_{21} - u_{21}v_{11},$$
$$\Delta_3 \equiv u_{11}v_{22} - u_{21}v_{12},$$
$$\Delta_4 \equiv u_{22}v_{11} - u_{12}v_{21},$$
$$\Delta_5 \equiv u_{22}v_{12} - u_{12}v_{22}.$$
Now we can transform the system (23) to the form:
$$\lambda_{i(j+1)} \Delta_1 \sqrt {j+1} =
- \lambda_{(i-1)j} \Delta_2 \sqrt {i} -
  \lambda_{i(j-1)} \Delta_3 \sqrt {j},$$
$$\lambda_{(i+1)j} \Delta_1 \sqrt {i+1} =
- \lambda_{(i-1)j} \Delta_4 \sqrt {i} -
  \lambda_{i(j-1)} \Delta_5 \sqrt {j} \eqno{(24)}$$
The system (23) is compatible if and only if $\Delta_2 = \Delta_5$
but this is exactly the condition that $[b_1,b_2]=0$ (see the
relation 1) (21)), i.e. this relation is always valid for the
canonical Bogolubov transformation. Under the unique condition
$\Delta_2 = \Delta_5$ we can obtain the general solution of the
system (23):
$$\lambda_{(2k)(2n+1)} = \lambda_{(2k+1)(2n)} = 0,$$
$$\lambda_{(2k)(2n)} = (-1)^{(n+k)} \sqrt {(2n)!}
\sqrt {(2k)!}\times$$
$$ \sum_{0 \leq s \leq n; s \leq k}\left [
\left ({\Delta_2 \over \Delta_1}\right )^{2s}
\left ({\Delta_3 \over 2 \Delta_1} \right )^{n-s}
\left ({\Delta_4 \over 2 \Delta_1} \right )^{k-s}
{1 \over (n-s)!(k-s)!(2s)!} \right ] \lambda_0,$$
$$\lambda_{(2k+1)(2n+1)} = (-1)^{(n+k+1)} \sqrt {(2n+1)!}
\sqrt {(2k+1)!} \times$$
$$\sum_{0 \leq s \leq n; s \leq k} \left  [
\left ({\Delta_2 \over \Delta_1}\right )^{2s+1}
\left ({\Delta_3 \over 2\Delta_1}\right )^{n-s}
\left ({\Delta_4 \over 2\Delta_1}\right )^{k-s}
{1 \over (n-s)!(k-s)!(2s+1)!} \right ] \lambda_0  \eqno{(25)}$$
Thus, for the vacuum state of Bose-fields $b$
we have:
$$|o>^{(b)} =$$ $$ \sum_{k,n} \left [ \lambda_{(2k)(2n)}|2k>_1^{(a)}
|2n>_2^{(a)} + \lambda_{(2k+1)(2n+1)}|2k+1>_1^{(a)}|2n+1>_2^{(a)}
\right ]$$

or in the operator form:
$$|0>^{(b)} = \lambda_0 e^{-{\Delta_4 \over 2\Delta_1}
(a_1^{\dagger})^2 -{\Delta_2 \over \Delta_1}a_1^{\dagger}
a_2^{\dagger} -{\Delta_3 \over 2\Delta_1}(a_2^{\dagger})^2}
|0>^{(a)} \eqno{(26)}$$
Thus, the operator
$$S_0(a^{\dagger})= \exp {\left( -{\Delta_4 \over 2\Delta_1}
(a_1^{\dagger})^2 -{\Delta_2 \over \Delta_1}a_1^{\dagger}
a_2^{\dagger} -{\Delta_3 \over 2\Delta_1}(a_2^{\dagger})^2\right )}
|0>^{(a)} \eqno{(27)}$$
is the operator of transformation of the vacuum states for the
most general two-mode Bogolubov transformations (20),(21).
The corresponding unitary "squeezed" operator U can be determined
now by recurrence procedure from the conditions
$$|n_1>_1^{(b)}|n_2>_2^{(b)} = U |n_1>_1^{(a)}|n_2>_2^{(a)}$$

\vskip .4in

\noindent{\bf References}
\vskip.2in
\noindent
\item{[1]} N.N.Bogolubov, J.Phys.USSR 11(1947), 23
\item{[2]} D.Stoler, Phys. Rev. D. 1(1970), 3217;
           D. 4(1971), 1925
\item{[3]} H.P.Yuen, Phys.Rev. A. 13(1976),2226
\item{[4]} R.Loudon, P.L.Knight, J.Modern Optics 34(1987), 709
\item{[5]} G.J.Milburn, J.Phys. A. 17(1984), 737
\item{[6]} C.M.Caves, B.L.Schumaker, Phys.Rev. D. 31(1985), 3068
\item{[7]} A.S.Davydov, Theory of solid. Nauka, Moscow. 1971
\item{[8]} A.V.Chizhov, R.G.Nazmitdinov, A.S.Shumovsky, Quant. Optics
           3(1991), 1

\bye